\documentclass{WileyMSP-template}
\usepackage{cite}
\usepackage{ragged2e}
\usepackage{soul}
\usepackage{color,xcolor}
\sethlcolor{pink} 
\begin{document}

\pagestyle{fancy}
\rhead{\includegraphics[width=2.5cm]{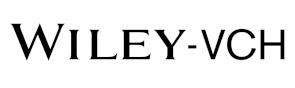}}

\title{Turnkey deterministic soliton crystal generation}

\maketitle


\author{Xinyu Yang\footnote[1]{Xiaotian Zhu and Xinyu Yang contributed equally to this work}}
\author{Xiaotian Zhu$^1$}
\author{Caitlin Murray,}
\author{Chawaphon Paryoonyong,}
\author{Xingyuan Xu,}
\author{Mengxi Tan*,}
\author{Roberto Morandotti,}
\author{Brent E. Little,}
\author{David J. Moss,}
\author{Sai T. Chu,}
\author{Bill Corcoran, and}
\author{Donglin Su*}


\begin{affiliations}
Miss. Xinyu Yang, Prof. Mengxi Tan, Prof. Donglin Su \\
Address: School of Electronic and Information Engineering, Beihang University, Beijing 100191, China\\
Email Address: xinyu$\_$yang1022@buaa.edu.cn, simtan@buaa.edu.cn, sdl@buaa.edu.cn\\

Miss. Xiaotian Zhu, Prof. Sai T. Chu\\
Department of Physics and Material Science, City University of Hong Kong, Tat Chee Avenue, Hong Kong, China\\
Email Address: xiaotizhu4-c@my.cityu.edu.hk, saitchu@cityu.edu.hk\\

Miss. Caitlin Murray, Dr. Chawaphon Paryoonyong, Prof. Bill Corcoran\\
Address: Department of Electrical and Computer System Engineering, Monash University, Clayton, VIC 3168, Australia.
ARC Centre of Excellence in Optical Microcombs for Breakthrough
Science, Australia.\\
Email Address: caitlin.murray1@monash.edu, park.prayoonyong@monash.edu, bill.corcoran@monash.edu\\
Prof. Xingyuan Xu\\
Address: State Key Laboratory of Information Photonics and Optical Communications, Beijing University of Posts and Telecommunications, Beijing 100876, China.\\
Email Address: xingyuanxu@bupt.edu.cn\\
Prof.Roberto Morandotti\\
Address: INRS-Énergie, Matériaux et Télécommunications, 1650 Boulevard Lionel-Boulet, Varennes, Québec J3X 1S2, Canada.\\
Email Address: roberto.morandotti@inrs.ca\\
Dr. Brent E. Little\\
Address: QXP Technologies Inc., Xi’an, China.\\
Email Address: brent.little@qxptech.com\\
Prof.David J. Moss\\
Address: Optical Sciences Centre, Swinburne University of Technology, Hawthorn, VIC 3122, Australia.
ARC Centre of Excellence in Optical Microcombs for Breakthrough Science, Australia.\\
Email Address:dmoss@swin.edu.au\\
 
\end{affiliations}



\justifying
\begin{abstract}

The deterministic generation of robust soliton comb has significant meaning for the optical frequency combs to be widely used in various applications. As a novel form of microcomb, Soliton crystal holds the advantages of easy generation, high conversion efficiency, and excellent thermal robustness. Here, we report the turnkey deterministic generation of ‘Palm-like’  soliton crystal with a free-running scheme. The robustness of the turnkey soliton crystal generation is also investigated in multiple aspects, including the success rate, the thermal robustness, and the long-term stability. The experiment results reveal our turnkey soliton crystal can achieve nearly a 100$\%$ success rate with a power variation less than ±1.5 dB over one hundred trials of two samples, is insensitive to thermal effect, and is robust to the environment during four-hour laboratory time.
\end{abstract}


\justifying
\section{Introduction}
It has been over two decades since octave spanning optical frequency combs (OFCs) were developed \cite{hansch2006nobel,hall2006nobel}, enabling absolute frequency referencing. Kerr microcombs \cite{del2007optical}– OFCs realized in micro-resonators – offer the possibility of realizing combs in a compact footprint, and the first reports\cite{moss2013new,razzari2010cmos,levy2010cmos} of fully integrated microcombs in CMOS compatible platforms triggered a tremendous wave of activity that has captured the imagination of the community since \cite{pasquazi2018micro,Kippenbergdks2018,gaeta2019photonic} and is still an extremely active field of research \cite{sun2024integrated,kudelin2024photonic,zhao2024all}. Beyond this, high repetition rates, high efficiencies and broad spectral coverage have made OFCs attractive for demonstrations focused on applications that span optical communications \cite{corcoran2024microcombs, marin2017microresonator,corcoran2020ultra}, frequency synthesis \cite{spencer2018optical, singh2020silicon}, quantum optics \cite{lu2019quantum,kues2019quantum,reimer2019high,kues2017chip,reimer2016generation}, microwave photonics \cite{xu2018advanced, wei2022measurement,sun2024integrated,kudelin2024photonic,zhao2024all,tan2021rf}, astronomical detection \cite{probst2020crucial}, and artificial intelligence \cite{xu202111, tan2023photonic, shastri2021photonics,feldmann2021parallel}. Among many different types of microcombs that have been reported, soliton microcombs are especially preferred for their high coherence and low noise \cite{wang2020advances}, and these properties are linked to solitons being self-reinforcing waves. There are various categories of soliton microcombs, such as bright dissipative Kerr solitons (DKSs) \cite{herr2016dissipative,herr2014temporal}, dark pulses \cite{lobanov2015frequency}, soliton crystals (SCs) \cite{cole2017soliton,corcoran2020ultra}, and laser-cavity solitons \cite{rowley2022self,bao2019laser}.

Bright DKS states were the earliest discovered soliton microcomb that could be generated in micro-ring resonators (MRRs) \cite{herr2014temporal,herr2016dissipative}, and operate in the red-detuned regime of MRRs having anomalous dispersion. However, when generating these soliton states, overcoming the fast resonance drift caused by the sudden decline of intracavity power that occurs at the onset of soliton generation, can be extremely challenging unless mitigated. Various control schemes have been applied to do this, such as fast frequency sweeping \cite{gaeta2019photonic}, power-kicking \cite{brasch2016photonic}, integrated thermal-tuning \cite{joshi2016thermally}, auxiliary laser pumping \cite{zhou2019soliton}, self-injection locking \cite{brasch2016bringing}, and forward and backward frequency sweeping \cite{guo2017universal}. Moreover, the typical internal conversion efficiency of DKSs is still only around 1\% \cite{bao2014nonlinear}.

Using a shifted resonant mode within a microresonator, either through coupled resonators or through mode interactions in a single resonator, enables higher efficiency microcomb generation \cite{xue2015mode,helgason2023surpassing}. These mode shifts can enable bright single solitons \cite{helgason2023surpassing}, dark pulses \cite{xue2015mode, fulop2018high} and SCs \cite{corcoran2020ultra}. Bright, single soliton states have been demonstrated recently with efficiencies up to 54\% \cite{helgason2023surpassing}, although this is achieved by reducing pump power after soliton generation, which means that the per-line power is not increased significantly. Dark pulse states have reached up to 50\%\cite{fulop2018high}, but lack the intrinsic self-reinforcing nature of bright states. Laser cavity soliton microcombs have achieved the highest efficiency of all microcombs to date, with the capability of reaching 100\% since they do not operate via the LLE and so do not require a CW background that limits the efficiency \cite{rowley2022self,bao2019laser}. They have also achieved self-starting and self recovering and operate by embedding a microring resonator in a nested cavity fiber loop configuration. 

SCs are self-organized ensembles of co-propagating solitons that fill the resonator in the angular domain, and can show efficiencies up to 40\% \cite{corcoran2020ultra}. Their generation is a function of pump power, pump wavelength, cavity dispersion, and modes shifted by an avoided mode crossing (AMX) which is characterized by its wavelength location and strength. For SCs, the AMX enables an extended background wave into the soliton waveform in the resonator, which forms an attractive soliton-soliton interaction, in turn enabling the ensemble of solitons to stably form \cite{wang2018robust,lu2021synthesized}. A key property of SCs is that there is little intracavity power change when forming the SC state, so the intracavity temperature (and hence optical the resonance undergoing pumping) does not have the significant shift that is associated with DKS states. Therefore, no complex pumping scheme is needed, and the robust SCs can be obtained by slow frequency sweeping or temperature tuning of the pump wavelength. This greatly simplifies the system required to generate the SC state, which can be experimentally obtained by slow frequency sweeping of the pump laser or temperature tuning of the resonator. 

Recent years have witnessed a surge in research on Kerr comb generations and applications with the aim moving microcombs out of the laboratory. This has been highlighted by advanced integrated photonics fabrication processes and packaging methods in order to greatly reduce the system size, weight, and power consumption (SWaP) as well as investigating the autonomous, deterministic, and robust generation of soliton combs. While deterministic generation has been shown for DKS states \cite{zhou2019soliton,zhang2024strong,kim2019turn,shen2020integrated,jin2021hertz,shu2022microcomb,dmitriev2022hybrid,ji2023engineered,weng2024turn}, other than for nested-cavity LCS microcombs, it has not been conclusively shown for high efficiency microcombs. 

Attempts to demonstrate deterministic, and robust generation of SCs have focused on so-called “perfect” soliton crystals (PSCs) that have been deterministically generated with a 100\% success rate \cite{karpov2019dynamics} by forward tuning with low pump power. Here, the mode shift is synthesized using an auxiliary laser \cite{lu2021synthesized}, to generate SCs with a mode spacing of 1 to 32 FSRs, when soliton number is over ten (490 GHz repetition rate), the success rate of PSC state generation is 100\%, and when soliton number is below 10, the success rate, while not 100\%, is still over 50\%, which reduces with decreasing soliton number. However, PSCs tend to have wide comb line spacings, or relatively low numbers of comb lines with significant power, which limits their use in many applications. Soliton crystals with defects, in contrast, can have comb lines at the resonator FSR, and many comb lines with significant power – leading to many different proof-of-concept demonstrations for a wide range of applications \cite{tan202211,cole2017soliton,advanced8606970,corcoran2020ultra,xuphtonicrf9096407}. These can be generated with a high likelihood of success, with \cite{wang2018robust} showing a probability of over 60\% for certain SC state. In \cite{mazoukh2024genetic}, the desired SCs, with line spacing of 48.9 to 150 GHz, can be generated by employing genetic algorithms as part of a closed-loop feedback control algorithm. However, these techniques for deterministic generation involve challenges such as complex setups resulting in wide mode spacings. Hence, while soliton crystals featuring defects look to be promising microcombs states, there has so far not been a simple, turn-key set up to generate these on demand. 

The pump-to-comb conversion efficiency (CE) of the soliton combs is also important for turnkey operation. Dark pulse turnkey generation using two coupled SiN ring resonators was previously reported with a CE of 41\% \cite{kim2019turn}. In addition, self-injection locking \cite{dmitriev2022hybrid} has achieved a CE of 40\% and 25\% for bright solitons with repetition rates of 3 THz and 1.2 THz respectively. This wide comb line spacing greatly reduces their range of applications. In contrast, bounded bright solitons have been generated in zero-dispersion rings with a CE of 26\% \cite{ji2023engineered}, but the comb spectrum led to flaws--either highly structured spectra featuring a broad bandwidth with a considerable power drop, or a relatively smooth spectrum with a less structured shape but with a much more limited bandwidth. To date, there have been no demonstrations repeatability generating the same soliton state simultaneously with stable comb line powers, together with turnkey operation.

In this work, we demonstrate deterministically generated turnkey ‘Palm-like’ 100 GHz (single FSR) SC microcombs using a free-running scheme. The turnkey SC generation is repeated 100 times using two different devices, both achieving a 100$\%$ success rate. Further, we repeat the experiments over a range of operating temperatures, still maintaining turnkey comb generation. Finally, we show that these states are robust to typical laboratory environmental variations, with line-by-line power variation lower than ±0.5 dB over the course of several hours, at the same time achieving a high efficiency of over 49\%. This is the first report of deterministic turnkey generation of robust SCs using an open-loop (no feedback) system, demonstrating perfect consistency. These results conclusively show that turnkey deterministic generation of ‘Palm-like’ SC combs, in different devices, at different temperatures, and robust over significant time frames can be achieved.  
\section{‘Palm-like' SC generation and simulation}
In this work we generate ‘Palm-like’ SCs in two different 100 GHz FSR 4-port (add-drop) MRRs fabricated on a complementary-metal- oxide-semiconductor (CMOS) compatible high-index doped silica glass platform. These ‘Palm-like’ states have supported a number of applications\cite{advanced8606970,corcoran2020ultra,tan202211,xuphtonicrf9096407,tan2023photonic}, demonstrating that these are very useful microcombs.  The radius of the MRR is 270 $\mu$m with a waveguide that is 3 $\mu$m wide and 1.5 $\mu$m thick. The device is pigtailed with a single mode fiber array to ensure the MRR is portable and reliably fiber coupled with very low loss – typically 1.0 - 1.5dB/facet . To help understand how ‘Palm-like’ SCs can be generated in these devices, the dispersion of the MRRs was measured to determine the AMX wavelength and strength. 

\begin{figure}[htbp!]
	\centering
	\includegraphics[width=0.8\textwidth]{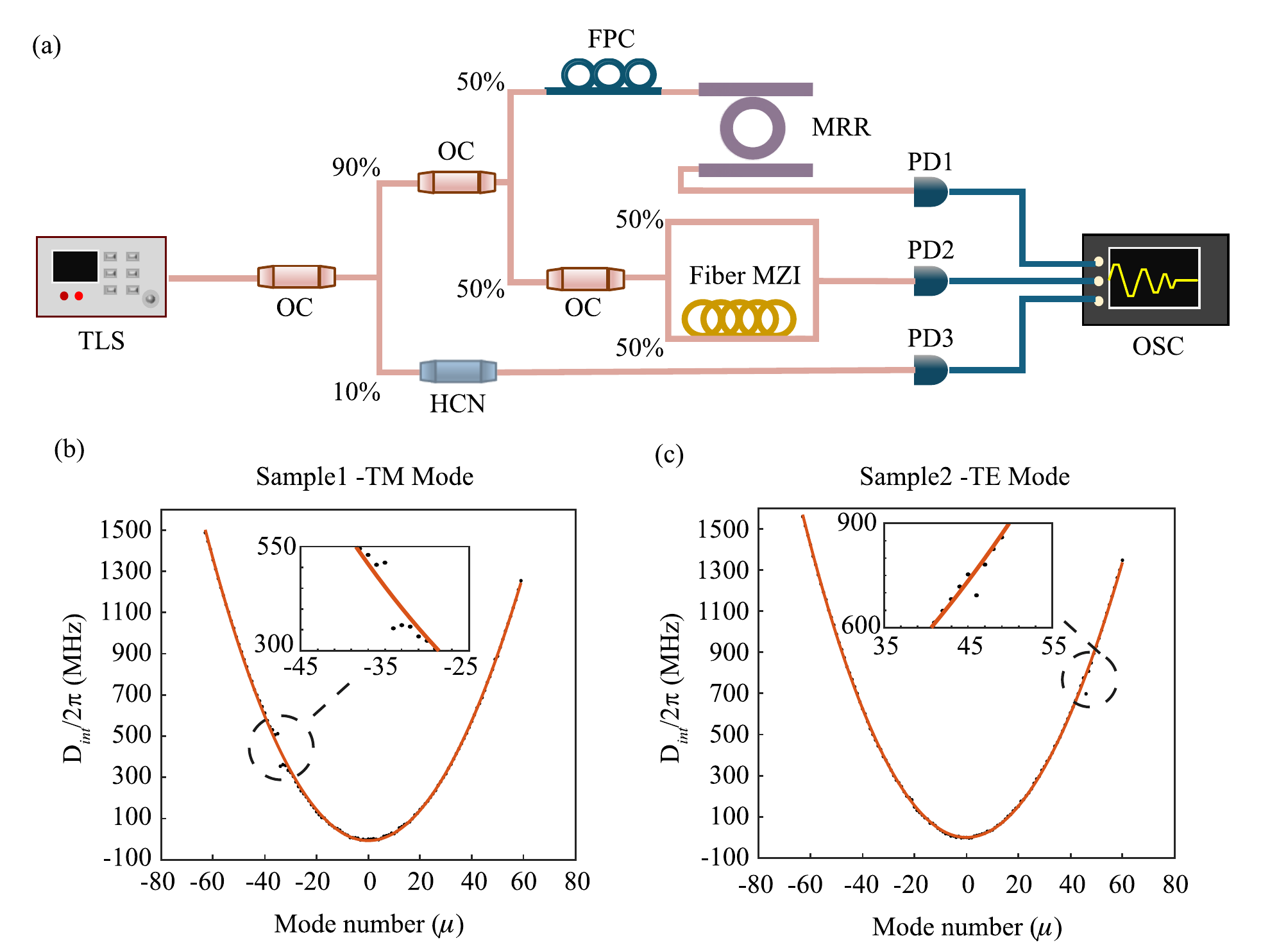}
	\captionsetup{font={stretch=1.5}}
	\caption[Dispersion]{The dispersion measurement setup and the results of 100 GHz MRR. TLS: tunable laser source; OC: optical coupler; FPC: fiber polarization controller; MZI: Mach-Zehnder interferometer; HCN: H$_{13}$C$_{14}$N calibration cell; PD: photodetector; OSC: oscilloscope. (a) The dispersion measurement setup; (b) The integrated dispersion of sample 1 with TM mode; (c) The integrated dispersion of sample 2 with TE mode.}
	\label{dispersion}
\end{figure}

The dispersion measurement setup is depicted in Fig.\ref{dispersion} (a). The tunable pump laser was swept across the MRR resonances as well as the Mach-Zehnder interferometer (MZI) and H$_{13}$C$_{14}$N (HCN) gas cell slowly at low power to eliminate effects of thermal drift on MRR resonances, with the oscilloscope triggered to record the traces when the pump laser started sweeping. The MZI acted as a reference to find the resonances with high resolution while the HCN cell was for the calibration of wavelengths. Using the data from MZI and HCN, the MRR resonant wavelengths could be accurately calculated using a Lorentzian resonance profile fit. The cavity modes ($\mu$) frequencies ($\omega$) are $\omega_{\mu}=\omega_{0}+{\mu}D_{1}+D_{int}$, where the $\omega_{0}$ is the center wavelength, $D_{1}/(2\pi)$ is the FSR of the cavity and the $D_{int}$ the integrated dispersion of the MRR which can be expressed as: $D_{int} =\omega -\omega _{0}-\cdots etc$. We chose the reference frequency ($\omega_{0}$) to match the frequency of the resonance closest to 1550 nm. In the absence of any AMXs, the $D_{int}$ is smooth and approximately parabolic. With the perturbation of the AMX, $D_{int}$ shows some discontinuities around the AMX location. The measured dispersion of the two rings used in the experiment is shown in Fig.\ref{dispersion} (b) and (c). The AMX strength (the largest difference between the $D_{int}$ of the modes shifted by the AMX) of sample 1 with the TM mode and sample 2 with TE mode is 131.78 MHz and 109.80 MHz respectively, and the corresponding AMX locations are 1522.435 nm and 1587.970 nm. As an approximate indication of pump wavelength, past experiments led us to link the AMX at 1522.435 nm of  the TM mode and 1587.970 nm of the TE mode, to the ability to generate a ‘Palm-like’ SC state if pumped close to 1550 nm and 1560 nm.   
 
\begin{figure}[htbp!]
	\centering
	\includegraphics[width=0.8\textwidth]{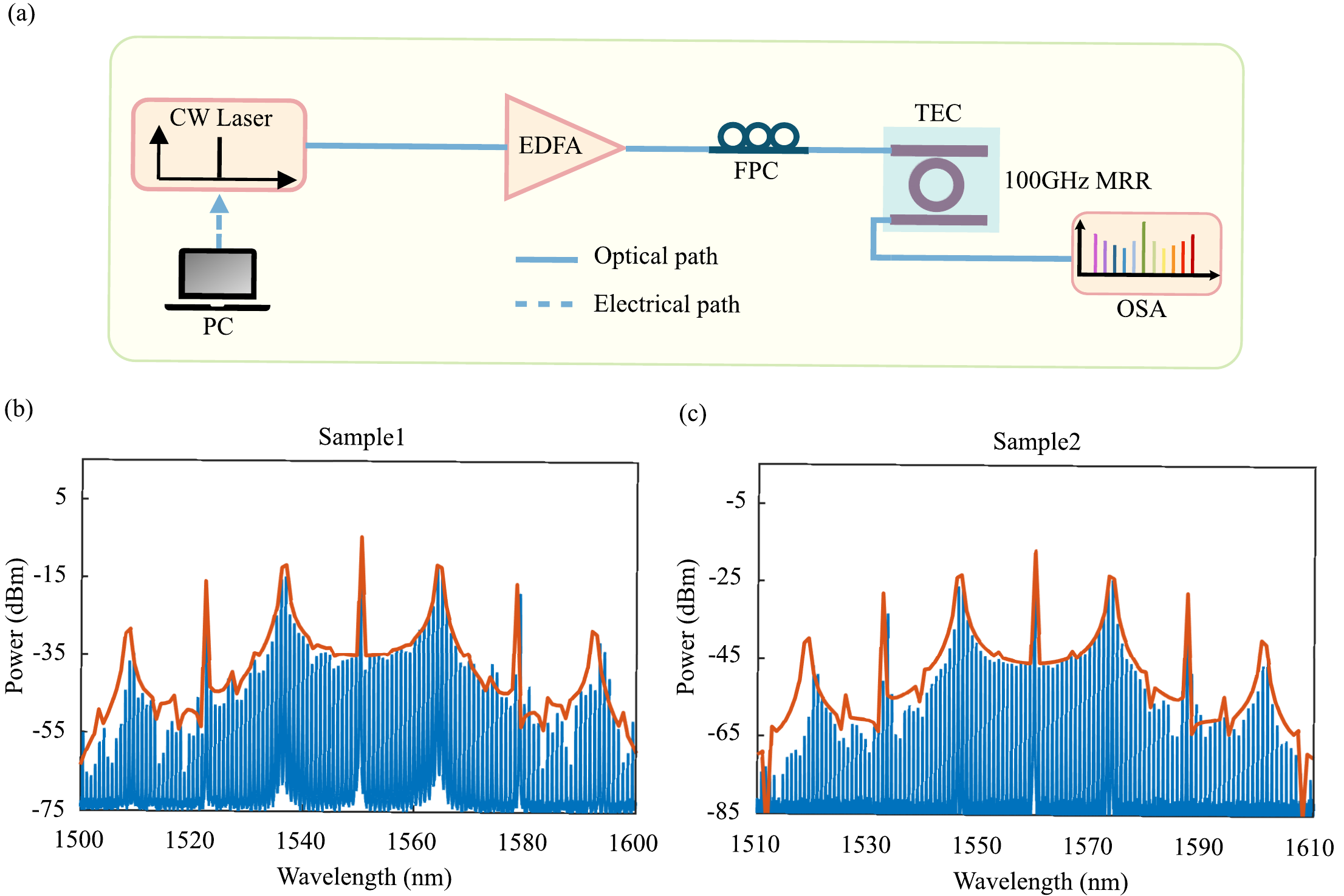}
	\captionsetup{font={stretch=1.5}}
	\caption[SC comb]{The generation setup and results of ‘Palm-like’ SC. EDFA: Erbium-doped fiber amplifier; TEC: thermo-electric cooler; OSA: optical spectrum analyzer; PC: personal computer. (a) The ‘Palm-like’ SC generation setup; (b) The ‘Palm-like’ SC generation and simulation results of sample 1;  (c) The ‘Palm-like’ SC generation and simulation results of sample 2.}
	\label{Fig2}
\end{figure}

The experimental setup for the ‘Palm-like’ SCs generation is shown in Fig.\ref{Fig2} (a). The device was mounted on a thermo-electric cooler (TEC) to stabilize the MRR resonances. The continuous wave (CW) tunable pump laser was connected to an Erbium-doped fiber amplifier (EDFA) directly to guarantee enough power (up to 2 W CW) for the SC generation. The output of EDFA was connected to a fiber polarization controller (FPC), which enabled the polarization of the incident light in the MRR to be set to the desired state. An attenuator was added at the drop port of MRR, then the light from the microcomb was passed into an optical spectrum analyzer (OSA). The TEC temperature was fixed at 25 $^{\circ}$C, and the wavelength of the CW laser manually tuned from shorter to longer wavelength until a ‘Palm-like’ SC was generated. When pumping the sample 1 (sample 2) with 950 mW (1.6 W) at 1550.33 (1559.76) nm, the desired ‘Palm-like’ state was generated, the experimental spectrum of two samples is depicted in Fig.\ref{Fig2} (b) and (c) with blue. Simulations were carried out using the normalized Lugiato-Lefever Equation (LLE) in Equation (\ref{equ1}) and the results are shown in Fig.\ref{Fig2} (b) and (c) with orange, with the simulation broadly replicating the features of the experimentally generated comb.
\begin{equation}
	\frac{\partial E(t,{\tau }' )}{\partial x}=\left [ -1+i(\left | E \right | ^{2}-\Delta )-2iD_{int}/k\right ]E+S  
	\label{equ1}
\end{equation}
where $\tau$ is the fast time corresponding to the azimuthal position in the ring, $\Delta$ is the detuning, $k=2\pi f_{0}/Q$  is the total loss rate, $E$ is the intracavity field and $S$ is the normalized input field.

\section{Deterministic turnkey generation}
\subsection{Repeatable turnkey SC generation}
\begin{figure}[htbp!]
	\centering
	\includegraphics[width=0.8\textwidth]{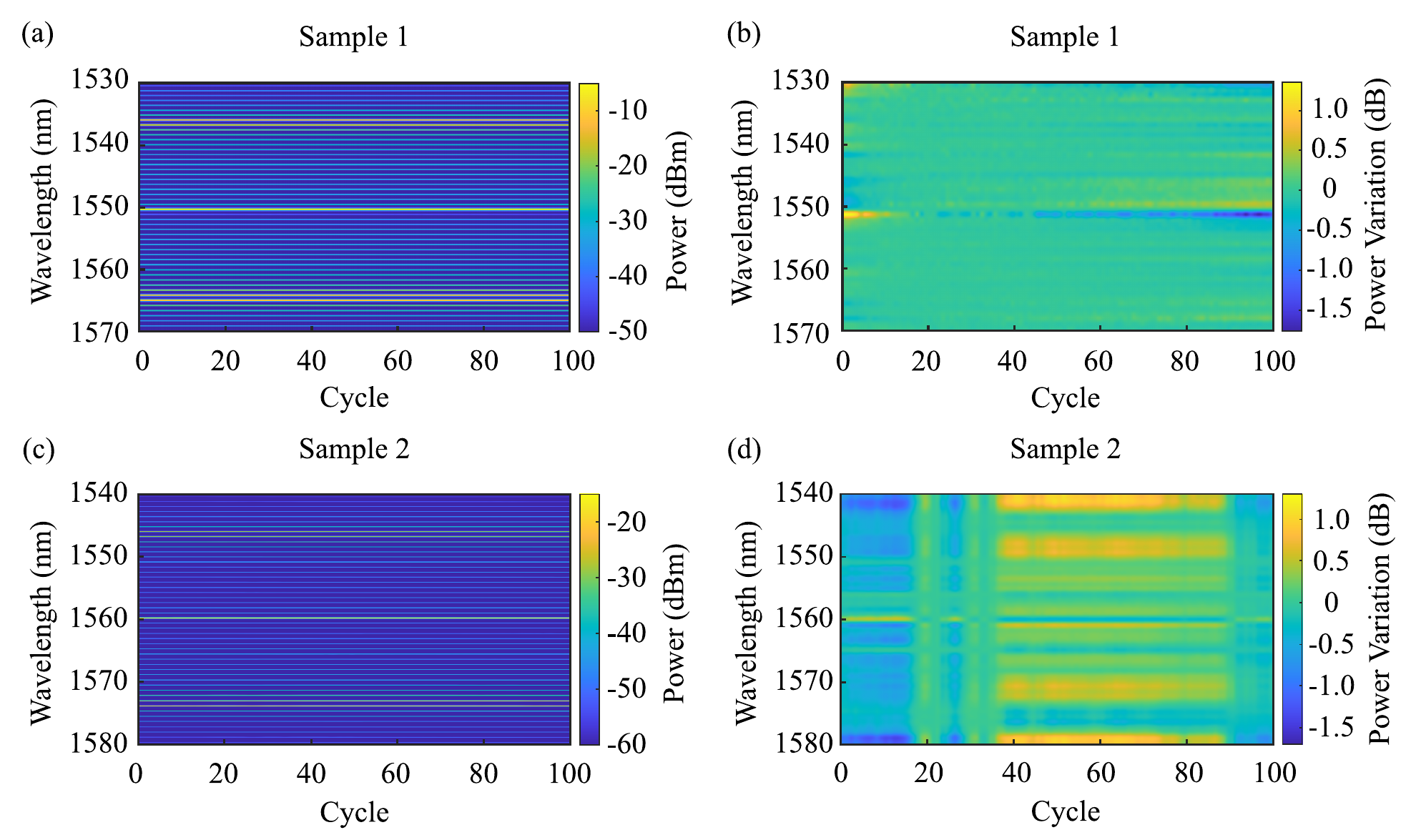}
	\captionsetup{font={stretch=1.5}}
	\caption[repeatable]{The repeatable turnkey ‘Palm-like’ SC generation. (a) The spectra evolution of 100 times deterministic turnkey ‘Palm-like’ SC generation  of sample 1; (b) The power variation of 100 times for sample 1; (c) The spectra evolution of 100 times deterministic turnkey ‘Palm-like’ SC generation  of sample 2; (d) The power variation of 100 times for sample 2. }
	\label{Fig3}
\end{figure}

Knowing that ‘Palm-like’ SCs could be generated at 1550.33 nm for sample 1 and 1559.76 nm for sample 2 using manually tuning, a program was then applied to control the laser for wavelength sweeping. To achieve turnkey generation, we set the start wavelength to 1540 nm and 1557 nm, then swept the laser wavelength to 1550.33 nm and 1559.76 nm for the two samples, respectively, resulting in the deterministic generation of the SC combs. To verify the repeatability of the turnkey SC generation, 100 cycles were executed, with the results showing that the same states were generated on completion of the laser sweep to the target wavelengths. The power distribution from 1530 nm to 1570 nm of sample 1 with 100 cycles is depicted in Fig.\ref{Fig3} (a), where the color bar shows the power spectrum density, and it can be seen that the power is ultra-stable over the 100 repeated experiments. Fig.\ref{Fig3} (b) shows the power variation map of resonances between 1530 and 1570 nm of sample 1 with 100 cycles, where the variation is shown with respect to the mid-value of the 100-cycle resonance power as a baseline, showing a variation within ±1.5 dB. Fig.\ref{Fig3} (c) shows the power distribution between 1540 and 1580 nm of sample 2 over 100 cycles and Fig.\ref{Fig3} (d) is the power variation, which is also within ±1.5 dB. The internal conversion efficiency (CE) was calculated for the two samples, defined as the ratio between all the comb lines power minus the pump power, to the total power. By analyzing the spectrum on OSA (1530-1570 nm for sample 1, 1540-1580 nm for sample 2), a CE for both MRRs was 49.07\% and 40.35\% respectively, the high value representing attractive performance for practical devices.

\begin{figure}[htbp!]
	\centering
	\includegraphics[width=0.8\textwidth]{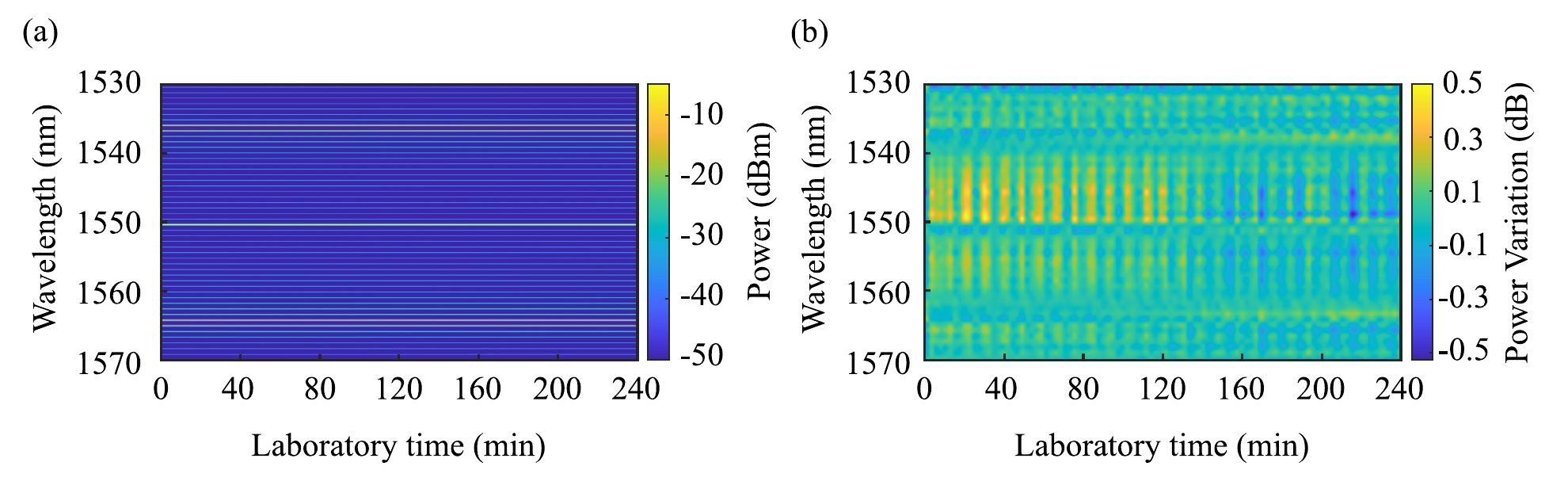}
	\captionsetup{font={stretch=1.5}}
	\caption[repeatable]{The long-term stability of SC. (a) The temporal evolution of the SC optical spectrum; (b) The power variation of the spectra within 4 hours.}
	\label{Fig5}
\end{figure}

In addition to the repeatability, the long-term stability is also an important factor that is essential for practical microcomb operation. Measurements of the microcomb spectra over 4 hours were performed at 25 $^{\circ}$C for sample 1-- the same conditions as for experiments described above, with the turnkey generated ‘Palm-like’ SCs spectra recorded every 3 minutes on an OSA. Fig.\ref{Fig5} (a)  shows the temporal evolution of a SC optical spectrum, where the color bar represents the optical power spectral density. Fig.\ref{Fig5} (b) illustrates the variation of power in the resonances over 4 hours, showing a variation within ±0.5 dB.   

\subsection{Turnkey SC characterization at different operating temperatures}
\begin{figure}[htbp!]
	\centering
	\includegraphics[width=0.8\textwidth]{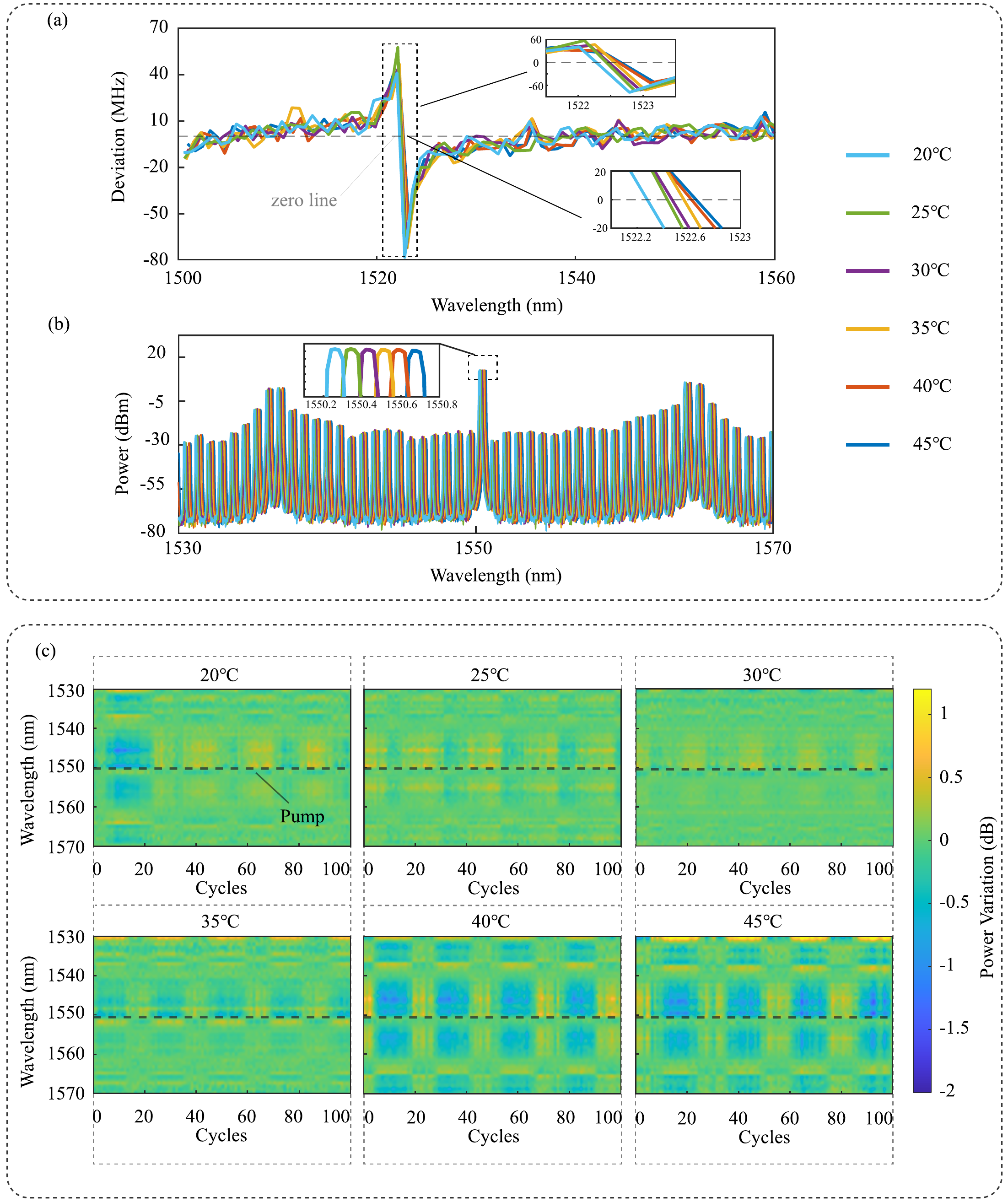}
	\captionsetup{font={stretch=1.5}}
	\caption[repeatable]{The repeatable turnkey ‘Palm-like’ SC generation with different temperatures. (a) The AMX measurement results with the temperature from 20 $^{\circ}$C to 45 $^{\circ}$C; (b) The generated SCs spectrum under different temperatures; (c) Power variation of the 100-cycle deterministic turnkey generation of ‘Palm-like’ SCs under 20 $^{\circ}$C, 25 $^{\circ}$C, 30 $^{\circ}$C, 35 $^{\circ}$C, 40 $^{\circ}$C and 45 $^{\circ}$C.}
	\label{Fig4}
\end{figure}
Temperature changes cause an effective index change in the MRR, further shifting the resonance and affecting the AMX property \cite{murray2023investigating}. To investigate whether thermal drifts affected the turnkey SC generation, a series of experiments were performed under a range of temperatures from 20 $^{\circ}$C to 45 $^{\circ}$C with a step size of 5 $^{\circ}$C. Here, only sample 1 was investigated since the two devices had performed similarly. To determine the thermal properties of the AMX, the dispersion was measured versus temperature \cite{murray2023investigating}. As shown in Fig.\ref{Fig4} (a) the AMX wavelength shifted about 0.38 nm and to longer wavelength when the temperature changed from 20 $^{\circ}$C to 45 $^{\circ}$C, representing a shift of much less than one FSR ($\sim$0.8 nm), with the change in AMX strengths between 12.37 MHz and 53.6 MHz. The pump wavelength was swept to generate ‘Palm-like’ SC at different temperature after characterizing the dispersion, with the generated spectra shown in Fig.\ref{Fig4} (b). As the pump wavelength was tuned to longer wavelength, with increasing temperature the calculated slope of the pump wavelength drift is 0.0169 nm/$^{\circ}$C. During the dispersion measurement, the resonance shifts were also measured simultaneously, being about 0.0166 nm/$^{\circ}$C and showing high consistency with the pump wavelength shift. These results reveal the small change in AMXs caused by temperature shifts did not affect the SC generation or the pump wavelength but rather can be predicted according to the thermal coefficient, similar to \cite{murray2023investigating}. To further explore the turnkey generation of ‘Palm-like’ SC at different temperatures, repeatable experiments as described in section 3.1 were performed. The initial wavelength was kept at 1540 nm with the final wavelengths being the target wavelength for each temperature. The repeatable turnkey experiments of the six temperatures are illustrated in Fig.\ref{Fig4} (c), and the spectra variation for different temperatures for the 100 repetitions was smooth while the pump wavelength displayed a minor drift.

\section{Conclusion}
In conclusion, we demonstrate turnkey generation of 100 GHz ‘Palm-like’ SC combs with open-loop control (no feedback) and investigate its repeatability and robustness. For 100 attempts at turnkey generation, with two different samples, we achieve a perfect 100$\%$ success rate in generating ‘Palm-like’ state, with line-by-line power variations over the entire 40 nm comb bandwidth being less than ±1.5 dB. The effect of temperature is investigated, showing that SC combs can be generated in a turnkey way over a wide range of operation temperatures. The temperature shifts have no influence on the deterministic generation of the SC combs and its repeatability. Moreover, in a standard lab environment over the course of a four-hour trial, line-by-line power variations are within ±0.5 dB, showing that the comb is stable under open-loop controls built into the commercial-off-the-shelf (COTS) laser and TEC. Our work shows that we achieved the deterministic generation of robust SC combs without any complex stabilizing schemes, which guarantees simple, stable, repeatable and predictable the comb generation. Deterministic generation of these soliton crystal combs opens up opportunities for these states to be used in applications where highly robust combs are desired, and where complex control loops would be problematic.      

\medskip
\textbf{Acknowledgements} \par 
This work is supported by the National Natural Science Foundation of China (No. 62293495) and by the Australian Research Council (ARC) Centre of Excellence in Optical Microcombs for Breakthrough Science “COMBS” (No. CE230100006) and by the Young Elite Scientists Sponsorship Program (No. 2023QNRC001). We thank Qihang Ai, Hanxiao Feng, and Zexu Wang for the contribution in the experiment setup.

\medskip

%
\bibliographystyle{MSP}
\bibliography{References}





\end{document}